\pdfoutput=1
\documentclass[11pt,twoside,a4paper,cmspaper,final,collab]{cms-tdr}
\def\svnVersion{8088d73}\def\svnDate{2026/06/15}\def\cmsCernNoTag{CERN-EP-2026-005}\def\cmsCernDate{\today}\def\cmsMessage{Published in Physical Review D as \href{http://dx.doi.org/10.1103/bwqs-vgqn}{\doi{10.1103/bwqs-vgqn}.}}
\begin{document}\cmsNoteHeader{B2G-24-006}

\newcommand{\EGfakes}{\ensuremath{\Pe \to \PGg}\xspace}
\newcommand{\EHfakes}{\ensuremath{\text{h} \to \PGg}\xspace}
\newcommand{\PQtst}{\HepParticle{\PQt}{}{*}\xspace}
\let\tstar\PQtst
\newcommand{\PAQtst}{\HepAntiParticle{\PQtst}{}{}\xspace}
\let\atstar\PAQtst
\newcommand{\tstartstar}{\ensuremath{\PQtst\PAQtst\xspace}}
\newcommand{\tstartg}{\ensuremath{\PQtst \to \PQt\Pg}\xspace}
\newcommand{\tstarta}{\ensuremath{\PQtst \to \PQt\PGg}\xspace}
\newcommand{\tstartz}{\ensuremath{\PQtst \to \PQt\PZ}\xspace}
\newcommand{\tg}{\ensuremath{\PQt\Pg}\xspace}
\newcommand{\ta}{\ensuremath{\PQt\PGg}\xspace}
\newcommand{\tstartstartgtg}{\ensuremath{\PQtst\PAQtst \to \PQt\PAQt\Pg\Pg}\xspace}
\newcommand{\tstartstartgta}{\ensuremath{\PQtst\PAQtst \to \PQt\PAQt\PGg\Pg}\xspace}
\newcommand{\pptstartstartgta}{\ensuremath{\Pp \Pp \to \PQtst\PAQtst \to \PQt\PAQt\PGg\Pg}\xspace}
\newcommand{\pptstartstartgtg}{\ensuremath{\Pp \Pp \to \PQtst\PAQtst \to \PQt\PAQt\Pg\Pg}\xspace}
\newcommand{\pptstartstar}{\ensuremath{\Pp \Pp \to \PQtst\PAQtst}\xspace}
\newcommand{\tgta}{\ensuremath{\PQt\PAQt\PGg\Pg}\xspace}
\newcommand{\tgtg}{\ensuremath{\PQt\PAQt\Pg\Pg}\xspace}
\newcommand{\mtstar}{\ensuremath{m_{\PQtst}}\xspace}
\newcommand{\ST}{\ensuremath{S_{\PQT}}\xspace}
\newcommand{\pp}{\ensuremath{\Pp\Pp}\xspace}
\newcommand{\Irel}{\ensuremath{I_{\text{rel}}}\xspace}
\newcommand{\BR}{\ensuremath{\mathcal{B}}\xspace}
\newcommand{\eptt}{\ensuremath{\epsilon_{\ttbar}}\xspace}

\newlength\cmsTabSkip\setlength{\cmsTabSkip}{1ex}
\newlength\cmsFigWidth
\ifthenelse{\boolean{cms@external}}{\setlength{\cmsFigWidth}{1\columnwidth}}{\setlength{\cmsFigWidth}{0.6\textwidth}}
\newcommand{\onehalf}{\ensuremath{1/2}\xspace}
\newcommand{\threehalf}{\ensuremath{3/2}\xspace}
\newcommand{\Gjets}{\ensuremath{\PGg\text{+jets}}\xspace}
\newcommand{\ttG}{\ensuremath{\ttbar\PGg}\xspace}
\newcommand{\sigmaii}{\ensuremath{\sigma_{i\eta i\eta }}\xspace}
\newcommand{\WG}{\ensuremath{\PW\PGg}\xspace}
\newcommand{\Wjets}{\ensuremath{\PW\text{+jets}}\xspace}
\newcommand{\ZG}{\ensuremath{\PZ\PGg}\xspace}
\newcommand{\Zjets}{\ensuremath{\PZ\text{+jets}}\xspace}
\newcommand{\msd}{\ensuremath{m_{\text{SD}}}\xspace}
\newcommand{\msdj}{\ensuremath{m^{{\text{j}}_{1}}_{\text{SD}}}\xspace}
\newcommand{\msdjj}{\ensuremath{m^{{\text{j}}_{2}}_{\text{SD}}}\xspace}
\newcommand{\sii}{\ensuremath{\sigma_{i\eta i\eta}}\xspace}
\newcommand{\Ich}{\ensuremath{I_{\text{chg}}}\xspace}
\newcommand{\Inu}{\ensuremath{I_{\text{nu}}}\xspace}
\newcommand{\IG}{\ensuremath{I_{\text{ph}}}\xspace}
\newcommand{\fr}{\ensuremath{f}\xspace}
\newcommand{\mgj}{\ensuremath{m_{\PGg\text{j}_{1}}}\xspace}
\newcommand{\pn}{\textsc{ParticleNet}\xspace}
\newcommand{\SohLimitE}{930\xspace}
\newcommand{\SohLimitO}{930\xspace}
\newcommand{\SthLimitE}{1390\xspace}
\newcommand{\SthLimitO}{1330\xspace}

\cmsNoteHeader{B2G-24-006}

\title{Search for pair production of heavy resonances in final states with a photon and large-radius jets in  proton-proton collisions  at \texorpdfstring{$\sqrt{s}=13\TeV$}{sqrt(s) = 13 TeV}}

\author*[cern]{A. Cern Person}
\date{\today}

\abstract{
A search for the pair production of heavy spin-\onehalf or spin-\threehalf resonances (\PQtst) in proton-proton collisions at $\sqrt{s}=13\TeV$ is presented.  Data collected with the CMS detector at the CERN LHC from 2016 to 2018 corresponding to an integrated luminosity of 138\fbinv are used. The analysis targets benchmark signal scenarios where one  \PQtst decays into a top quark (\PQt) and a photon (\PGg), and the other into a \PQt quark and a gluon (\Pg), \ie, $\Pp\Pp\to\PQtst\PAQtst\to\PQt\PQt\PGg\Pg$. All-hadronic final states from the \PQt pair decay chain are selected using jet substructure techniques.  The signal is probed as a function of the  \PQtst candidate mass, which is reconstructed using the photon and a top quark candidate jet. No significant deviation from the background-only hypothesis is found.  Observed (expected) upper limits on the signal cross section at 95\% confidence level are set, excluding masses of spin-\onehalf  \PQtst particles below 930\,(930)\GeV and spin-\threehalf \PQtst particles below 1330\,(1390)\GeV. This analysis marks the first search for heavy resonances in the $\ttbar\PGg\Pg$ channel.  Exploiting the high-energy photon to reduce the backgrounds, this search achieves sensitivity competitive with $\Pp\Pp\to\PQtst\PAQtst\to\ttbar\Pg\Pg$ searches for spin-\onehalf \PQtst despite the small expected $\PQtst\to\PQt\PGg$ branching fraction.}

\hypersetup{%
pdfauthor={CMS Collaboration},%
pdftitle={Search for pair production of heavy resonances in final states with a photon and large-radius jets in  proton-proton collisions  at sqrt(s) = 13 TeV},%
pdfsubject={CMS},%
pdfkeywords={CMS, top}}%
\maketitle 

\section{Introduction}
\label{sec:introduction}

The standard model (SM) of particle physics has been successful in describing a large number of observations. 
However, it leaves several fundamental questions unanswered.
For example, the SM cannot explain the vast disparity between the strengths of the electroweak (EW) force and gravity~\cite{Arkani-Hamed:1998jmv}. Additionally, the SM fails to elucidate the nature of dark matter. 
Beyond-the-SM theories, including those with extra spatial dimensions~\cite{Hewett:1988xc} and composite Higgs scenarios~\cite{Vignaroli:2012sf}, can address these open questions. Some of these models predict that the top quark may have excited states  or have vector-like quark partners.

For a top quark vector-like  partner (\PQT) with an electric charge of  $+2/3$, the dominant decay modes are expected to be the tree-level processes $\PQT\to \Pb \PW $, $\PQT\to \PQt \PZ $, and $\PQT\to \PQt \PH $~\cite{DeSimone:2012fs}. 
Both the ATLAS and CMS experiments at the CERN LHC have searched for these signatures, placing lower constraints on the mass of the \PQT of up to  1.42\TeV at 95\% confidence level (\CL), depending on the \PQT branching fractions (\BR)~\cite{CMS_tH_tZ_single, CMS_tH_tZ_pair, CMS_bW_pair, CMS:2024qdd, CMS_tZ_single, ATLAS_tZ_single, ATLAS_tH_tZ_bW_pair, ATLAS_tZ_pair, CMS:2023agg}. These  decay modes  could be suppressed  if the mixing angle, which governs the interaction strength between \PQT and W, Z, or H, is small. Consequently, the loop-induced decays $\PQT\to\tg$ and $\PQT\to\ta$ can dominate~\cite{Serra:2015xfa,Bizot:2018tds,Han:2018hcu,Benbrik:2019zdp}.  These decays  are similar to those of an excited top quark (\tstar), which proceed via dipole operators~\cite{Baur:1989kv}. Among these, the decay \tstartg is expected to be dominant because of the strong coupling, while \tstarta would be suppressed by the electromagnetic coupling. 
Searches for \tstar pair production, assuming the \tstartg decay mode, have been performed by the CMS Collaboration using proton-proton ($\Pp\Pp$) collisions at center-of-mass energies of 8 and 13\TeV. These studies  resulted in a 1050 (1700)\GeV 95\% \CL lower mass limit for a spin-\onehalf (spin-\threehalf) \tstar, assuming $\BR(\tstartg)=100$\%~\cite{CMS:2013tlq,CMS:2017ixp,CMS:2024zpb}. 
Despite the expected smallness of  $\BR(\tstarta)$ with respect to $\BR(\tstartg)$, the sensitivity of the \tstartstartgta  channel could be similar to that of the \tgtg channel  because of the smaller backgrounds associated with its photon signature~\cite{Alhazmi:2018whk}. 

The interactions of spin-\onehalf \tstar with SM particles are described by a dipole-moment-like interaction between the t quark and \tstar~\cite{Alhazmi:2018whk}. The interactions of a spin-\threehalf \tstar are described by the Rarita--Schwinger Lagrangian~\cite{Rarita:1941mf}.
Although the spin-\onehalf and spin-\threehalf \tstar are both assumed to be color-triplet particles, the $\Pp\Pp\to\tstartstar$ cross sections are not the same because of the spin-dependent couplings to SM gluon fields. 
For example, the production cross section for a pair of spin-\onehalf \tstar 
with a mass of 1\TeV is $\approx$26\unit{fb}, whereas for a pair of spin-\threehalf $\tstar$ 
particles with the same mass it is $\approx$260\unit{fb}.
The branching fractions of the \tstar depend on the coupling strength to gluons and photons.  Simple coupling ratio arguments suggest absolute branching fractions of $\BR(\tstarta)= 3\% $ and $\BR(\tstartg) = 97\% $. These and other theoretical assumptions adopted in this analysis follow Ref.~\cite{Alhazmi:2018whk}.

This paper describes the first search at the LHC for pair-produced \tstar with spin-\onehalf or spin-\threehalf in the  \tgta channel in pp collisions at $\sqrt{s}=13\TeV$.
The data used in the analysis correspond to an integrated luminosity of 138\fbinv, collected by the CMS experiment at the LHC during 2016--2018. The large expected mass of the \tstar implies that the partons produced in its decay will have very large momenta, resulting in a photon and top quark of high transverse momentum (\pt). Because of the large Lorentz boost of the top quark, its decay products are typically expected to be merged into one large radius particle jet.  Thus, events characterized by an isolated photon with high-\pt and multiple large-radius jets are used in this analysis.

\begin{figure}[h!t]
\centering
      \includegraphics[width=\cmsFigWidth]{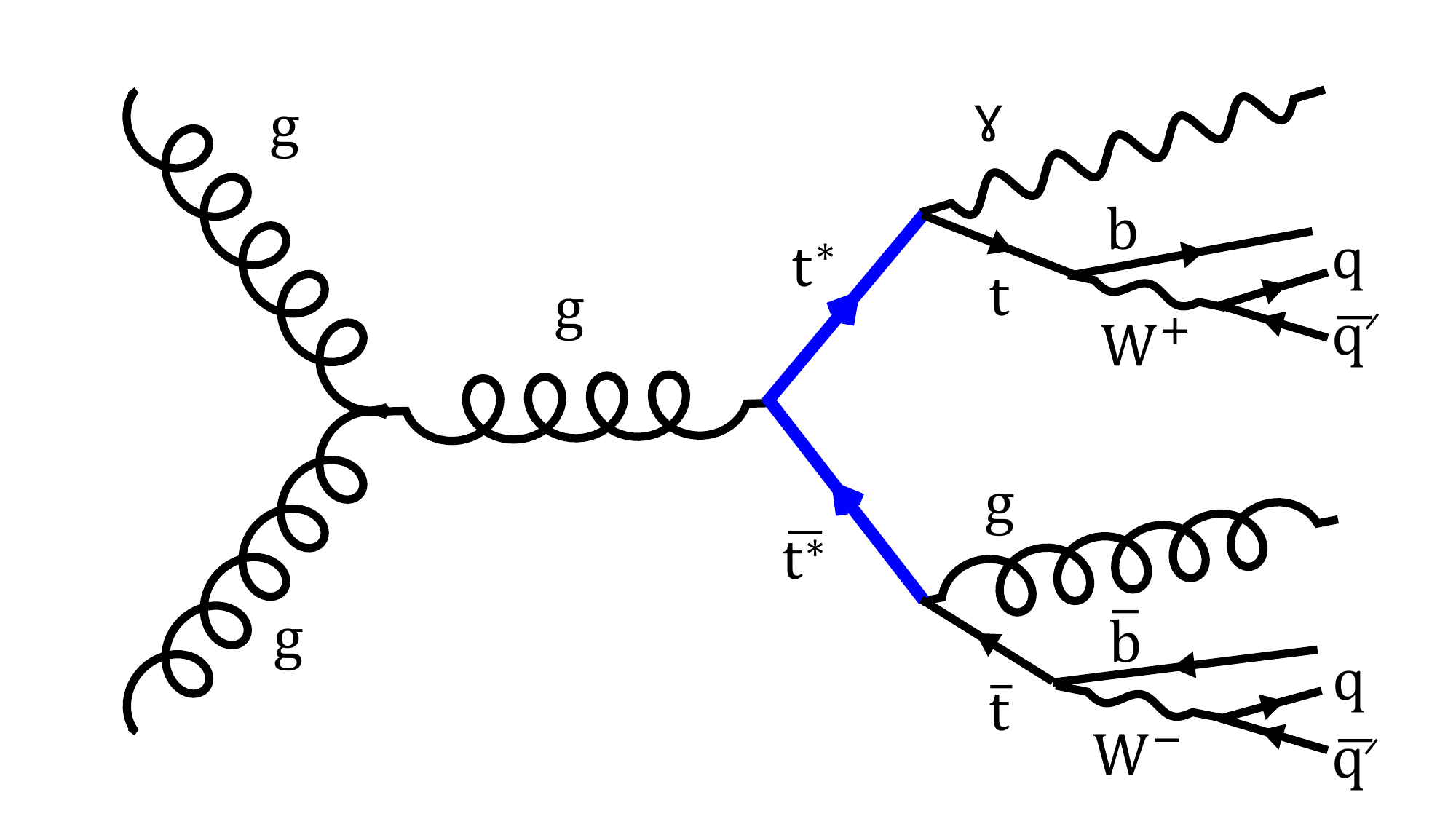}
    \caption{Representative LO Feynman diagram for pair production of \tstar in the \tgta channel. 
    \label{FeynmanDiagram}}
\end{figure}

This search utilizes a state-of-the-art machine-learning algorithm, known as the \pn tagger, to identify the large-radius jets as originating from top quarks~\cite{Qu:2019gqs}.
In this study, we employ the reconstruction of the mass of a single \tstar candidate, using a photon and a top quark  jet, to search for an excess of events in the mass distribution that would correspond to the presence of a \tstar.
The corresponding leading-order (LO) Feynman diagram for the search channel targeted in this paper is shown in Fig.~\ref{FeynmanDiagram}. 

\section{The CMS detector}
\label{sec:CD}

The CMS apparatus is a multipurpose, nearly hermetic detector, designed to trigger on~\cite{CMS:2020cmk,CMS:2016ngn,CMS:2024aqx} and identify electrons, muons, photons, and (charged and neutral) hadrons~\cite{CMS:2020uim,CMS:2018rym,CMS:2014pgm}. Its central feature is a superconducting solenoid of 6\unit{m} internal diameter, providing a magnetic field of 3.8\unit{T}. Within the solenoid volume are a silicon pixel and strip tracker, a lead tungstate crystal electromagnetic calorimeter (ECAL), and a brass and scintillator hadron calorimeter (HCAL), each composed of a barrel and two endcap sections. Forward calorimeters extend the pseudorapidity ($\eta$) coverage provided by the barrel and endcap detectors. Muons are reconstructed using gas-ionization detectors embedded in the steel flux-return yoke outside the solenoid. More detailed descriptions of the CMS detector, together with a definition of the coordinate system used and the relevant kinematic variables, are reported in Refs.~\cite{CMS:2008xjf,CMS:2023gfb}.

\section{Data and simulated event samples}
Events of interest are selected using a two-tiered trigger system. The first level, 
composed of custom hardware processors, uses information from the calorimeters and muon detectors to select 
events at a rate of around 100\unit{kHz} within a fixed latency of about 4\mus~\cite{CMS:2020cmk}. 
The second level, known as the high-level trigger, consists of a farm of processors running a version of the 
full event reconstruction software optimized for fast processing, and reduces the event rate to 
around 1\unit{kHz} before data storage~\cite{CMS:2016ngn}.
Collision events are selected using a high-\pt photon trigger. The \pt thresholds for the photon trigger are 175\GeV for data collected during 2016, and 200\GeV for data collected in 2017 and 2018, reflecting adjustments to account for the changing LHC luminosity conditions.
The trigger efficiencies, measured in $\PZ\to\Pe\Pe$ samples, are  greater than 99\% for photon with $\pt>240$\GeV~\cite{CMS:2021zxu}.

Monte Carlo (MC) simulated events are used to  model  most of the background processes and the \tstar signals. 
The next-to-next-to-leading order (NNLO) NNPDF3.1~\cite{Ball:2017nwa} parton distribution function (PDF) set is used in event production.
Parton showering and hadronization are performed with \PYTHIA v8.240~\cite{Sjostrand:2014zea} using the underlying event tune  CP5~\cite{Sirunyan:2019dfx}.
All the MC generated samples are passed through the full simulation of the CMS detector based on \GEANTfour~\cite{Agostinelli:2002hh}.
Simulated samples include additional $\Pp\Pp$ interactions occurring within the same or nearby bunch crossings (pileup). Events in these samples are reweighted so that the pileup distribution matches that of the data.

Signal samples for the pair production of spin-\onehalf and spin-\threehalf \tstar along with their decays, $\Pp\Pp\to\tstartstartgta$, are generated at LO with the \MGvATNLO v.2.6.5 generator~\cite{Alwall:2014hca}.  
The narrow-width approximation is assumed to be valid in the mass range considered in this analysis~\cite{Alhazmi:2018whk}.
In practice, the signal samples are generated with a fixed \tstar width of 10\GeV, which is much smaller than the experimental resolution of the reconstructed \tstar mass (\mtstar). Under this approximation, signal samples are produced for \mtstar values ranging from 700\GeV to 2\TeV in 100\GeV steps, with four additional mass points at 2.25, 2.5, 2.75, and 3\TeV.
All signal samples are normalized to the cross section of \tstartstar production, calculated at LO, times the product of $\mathcal{B}(\tstartg)=0.97$ and $\mathcal{B}(\tstarta)=0.03$~\cite{Alhazmi:2018whk}.

The background processes considered in this analysis include several key contributions. These encompass SM top quark-antiquark pair production (\ttbar), as well as \ttbar produced in association with a photon (\ttG). Additionally, single top quark production occuring through the $t$-channel, tW-channel (associated with a W boson), and $s$-channel is considered. Other important backgrounds include multijet production and processes involving the production of a photon, Z, or W boson in association with jets (\Gjets, \Zjets, and \Wjets). Processes where a  Z or a  W boson is produced alongside a photon (\ZG  and  \WG ) are also included.
The SM \ttbar, and single top quark production processes in the $t$-channel and tW-channel are simulated with  POWHEG v2.0 at next-to-leading order (NLO)~\cite{bib:powheg0,bib:powheg2,Frixione:2007vw}.
The \MGvATNLO v.2.6.5 generator is used to simulate other processes.
The multijets, \Gjets, \Wjets, \WG, and \ttG samples are generated at LO; other MC samples are generated at NLO. Jets produced at the matrix element level are matched with those generated by \PYTHIA parton shower using the MLM and FxFx merging schemes for LO and NLO samples, respectively~\cite{Frederix:2012ps}. Possible overlaps between events with a photon generated at the matrix element level in the \ttG,  \Gjets, \ZG, and \WG processes and events with a photon generated by \PYTHIA at the showering step in the \ttbar, multijet, \Zjets, and \Wjets processes are removed~\cite{CMS:2021klw}.

\section{Event reconstruction}
\label{sec:ER}
The global event description based on the particle-flow (PF) algorithm~\cite{CMS:2017yfk} aims to reconstruct and identify each individual particle in an event, with an optimized combination of all subdetector information. In this process, the identification of the particle type (photon, electron, muon, charged or neutral hadron) plays an important role in the determination of the particle direction and energy. Photons 
are identified as ECAL energy clusters not linked to the extrapolation of any  trajectory of a charged particle originating from the primary vertex. Electrons are identified as a primary charged-particle track and potentially multiple ECAL energy clusters corresponding to this track extrapolation to the ECAL and to possible bremsstrahlung photons emitted along the way through the tracker material. Muons are identified as tracks in the central tracker consistent with either a track or several hits in the muon system, and associated with energy deposits in calorimeters compatible with the muon hypothesis. Charged-hadrons are identified as charged-particle tracks neither identified as electrons, nor as muons. Finally, neutral hadrons are identified as HCAL energy clusters not linked to any charged-hadron trajectory, or as a combined ECAL and HCAL energy excess with respect to the expected charged-hadron energy deposit.

The primary vertex (PV) is taken to be the vertex corresponding to the highest-momentum scattering in the event, evaluated using tracking information alone, as described in Section 9.4.1 of Ref.~\cite{CMS-TDR-15-02}.
The PV is  used to associate the reconstructed objects in the event, such as jets and photons, and to calculate kinematic variables such as \pt and $\eta$ relative to the PV.

The energy of photons is obtained from the ECAL measurement. The energy of electrons is determined from a combination of the track momentum at the main interaction vertex, the corresponding ECAL cluster energy, and the energy sum of all bremsstrahlung photons attached to the track. The energy of muons is obtained from the corresponding track  curvature. The energy of charged hadrons is determined from a combination of the track momentum and the corresponding ECAL and HCAL energies, corrected for the response function of the calorimeters to hadronic showers. Finally, the energy of neutral hadrons is obtained from the corresponding corrected ECAL and HCAL energies.

In the barrel section of the ECAL, an energy resolution of about 1\% is achieved for unconverted or late-converting photons in the tens of GeV energy range. The energy resolution of the remaining barrel photons is about 1.3\% up to $\abs{\eta} = 1$, changing to about 2.5\% at $\abs{\eta} = 1.4$~\cite{CMS:2020uim}. 
A standard identification (ID) working point, called ``medium ID'' in Ref.~\cite{CMS:2020uim}, with an average efficiency of 80\% for selecting genuine photons is used in this analysis. 
The photon  ID is based on requirements on the ratio
of the energy deposited in the closest HCAL tower to
that of the ECAL and shower profile of the photon ($\sii$). 
The \sii  observable, defined as the energy-weighted root-mean-square of the electromagnetic shower profile along the $\eta$ direction, quantifies the lateral spread of the shower in units of the ECAL crystal size~\cite{CMS:2015myp}.
Selected photons must also
satisfy the photon isolation requirement based
on the scalar \pt sum of charged hadron ($\Ich$), neutral hadron, and additional photon candidates reconstructed by the PF algorithm within a cone of $\Delta R = \sqrt{\smash[b]{(\Delta \eta)^2 + (\Delta \phi)^2}} <0.3$ around its axis~\cite{CMS:2020uim}, where $\phi$ is azimuthal angle in radians. 
The photon is distinguished from identified electrons in the detector by requiring no hits in the inner tracker layers close to the photon direction. 
The photon ID and isolation efficiencies are measured separately in data and simulation with $\PZ\to\Pe\Pe$ events, with the electron tracks removed to emulate photon-like ECAL superclusters except in cases where photon conversions occur.
Scale factors (SF) are applied to simulated events to correct for differences between the behavior of reconstructed photons in the simulation and those observed in collision data.
The same $\PZ\to\Pe\Pe$ events are utilized to calibrate the photon energy scale and resolution~\cite{CMS:2020uim}.

In each event, hadronic jets are formed by clustering the reconstructed PF candidates using the infrared- and collinear-safe anti-\kt algorithm~\cite{Cacciari:2008gp}, as implemented in the \FASTJET package~\cite{Cacciari:2011ma}.
Large-radius jets (''AK8 jets''), with a distance parameter of 0.8, are used to reconstruct hadronically decaying, highly Lorentz-boosted top quark candidates. The jet momentum is determined as the vectorial sum of the jet constituents, and is found from simulation to be, on average, within 5 to 10\% of the true momentum over the entire \pt spectrum and detector acceptance~\cite{CMS:2016lmd}. Pileup interactions can contribute additional tracks and calorimetric energy depositions to the jet momentum. The pileup-per-particle identification algorithm~\cite{Sirunyan:2020foa,Bertolini:2014bba} is used to mitigate the effect of pileup at the reconstructed-particle level, making use of local shape information, event pileup properties, and tracking information. A local shape variable is defined, which distinguishes between collinear and soft diffuse distributions of other particles surrounding the particle under consideration. The former is attributed to particles originating from the hard scatter and the latter to particles originating from pileup interactions. Charged particles identified to be originating from pileup vertices are discarded. For each neutral particle, a local shape variable is computed using the surrounding charged particles compatible with the PV within the tracker acceptance ($\abs{\eta} < 2.5$), and using both charged and neutral particles in the region outside of the tracker coverage. The momenta of the neutral particles are then rescaled according to their probability to originate from the PV deduced from the local shape variable, superseding the need for jet-based pileup corrections~\cite{Sirunyan:2020foa}. 

Jet energy corrections are derived from simulation studies so that the average measured energy of jets becomes identical to that of particle-level jets. In situ measurements of the momentum balance in dijet, $\PGg + \text{jet}$, $\PZ + \text{jet}$, and multijet events are used to determine any residual differences between the jet energy scale in data and in simulation, and appropriate corrections are made~\cite{CMS:2016lmd}. Additional selection criteria are applied to each jet to remove those potentially dominated by instrumental effects or reconstruction failures~\cite{Sirunyan:2020foa}.
The groomed jet mass is calculated after applying a modified mass-drop algorithm~\cite{Dasgupta:2013ihk,Butterworth:2008iy}, known as the soft-drop mass (\msd)~\cite{Larkoski:2014wba}, to AK8 jets with  parameters $\beta=0$, $z_\text{cut}=0.1$, and $R_0 = 0.8$. 

Hadronic decays of top quarks are identified using the \pn tagger, a graph neural network based jet tagging algorithm~\cite{Qu:2019gqs}. The PF candidates and secondary vertices associated with the AK8 jets are used as input to the algorithm, which is designed to identify different jet types. In this search, the discrimination of \PQt jets against other jet types is utilized.
The AK8 jets with a \pn tagger score above a specific threshold are classified as ``\PQt-tagged'' jets.
For the selected thresholds on the \pn score, the efficiency for identifying AK8 jets originating from the hadronic decays of top quarks within the kinematic range $ 400 < \pt < 600 $\GeV  (where \pt denotes the transverse momentum of the AK8 jet)  is approximately 82\%. Meanwhile, the misidentification rate for jets originating from the hadronization of gluons or light-flavor quarks (\PQu, \PQd, \PQs) is about 0.1\%.
 The \msd and \PQt tagging SFs  are derived using \ttbar events where one top quark decays into a lepton, neutrino, and a b jet, while the other decays to three jets.
 
Isolated muon and electron candidates are reconstructed to veto events containing them. Muons are identified using dedicated selection criteria for high-momentum muons, as described in Ref.~\cite{CMS:2019ied}. The electron identification is performed using a multivariate analysis discriminant that combines various measured properties from the tracker and ECAL~\cite{CMS:2015xaf}.  Muons (electrons) must satisfy $\abs{\eta}<2.4$ and $\pt>55~(40)~\GeV$. Electrons in the  gap between the ECAL barrel and endcap calorimeters ($1.44<\abs{\eta}<1.56$) are excluded, because the reconstruction of an electron object in this region is not optimal. Selected leptons must be separated from the photons and jets by $\Delta R>0.4$ and 0.8, respectively. 

\section{Event selection and analysis strategy}
\label{sec:evsel}

The analysis selection requires exactly one photon in the ECAL barrel region, i.e., $\abs{\eta}< 1.44$, with $\pt>240$\GeV. Events with photons reconstructed in the endcap regions are excluded due to negligible signal sensitivity. Photons are required to be separated from AK8 jets by $\Delta R >0.8$.  
Events containing muons or electrons are excluded, as described above.
The AK8 jets are required to be within the region $\abs{\eta}< 2.4$ and must have $\pt>300$\GeV. 
Signal events may contain three AK8 jets originating from the top quark, top antiquark, and gluon produced in \tstartstar decays, provided all jets fall within the defined acceptance. 
Events in the signal region (SR) are required to contain at least two AK8 jets, of which one (j$_1$) must be \PQt tagged with $125 < \msdj < 225~\mathrm{GeV}$ and another (j$_2$) must satisfy $\msdjj > 50~\mathrm{GeV}$.
No additional requirement is imposed on the total number of AK8 jets in the event.
With these selection criteria, most background events that do not contain a top quark are removed. Consequently, additional signal characteristics, such as the presence of high-\pt jets from gluons or b quarks, are not utilized. This approach maintains  generality and preserves high signal  efficiency. 
The  efficiency for the spin-\threehalf signal is up to 5\% higher than that of the spin-\onehalf case,  in the mass range 700--2000\GeV, with absolute efficiencies ranging from approximately 8\% to 15\%. This difference arises because spin-\threehalf \tstar events are produced more centrally than those of spin-\onehalf.

We search for the top quarks from $\tstar$ decays by requiring at least one \PQt-tagged AK8 jet. The probability of successfully reconstructing both \PQt candidates is less than 20\% for $\tstar$ particles with masses below 2\TeV in the SR.
In events with one \PQt-tagged AK8 jet,  there is an ambiguity in the $\tstar$ reconstruction since the selected top quark and photon candidates can originate from different \tstar parents. 
For events with two \PQt-tagged AK8 jets, the \PQt-tagged jet with the lower \msd is chosen to associate to the photon and reconstruct the $\tstar$ mass (\mgj). This approach achieves a correct pairing efficiency between the \PQt and \PGg candidates of about 50--60\% for $\mtstar<2\TeV$. Importantly, the effectiveness of this assignment choice  is largely independent of the $\tstar$ mass and spin, ensuring consistency across different signal scenarios. 

\begin{figure}[tb]
\centering
\includegraphics[width=\cmsFigWidth]{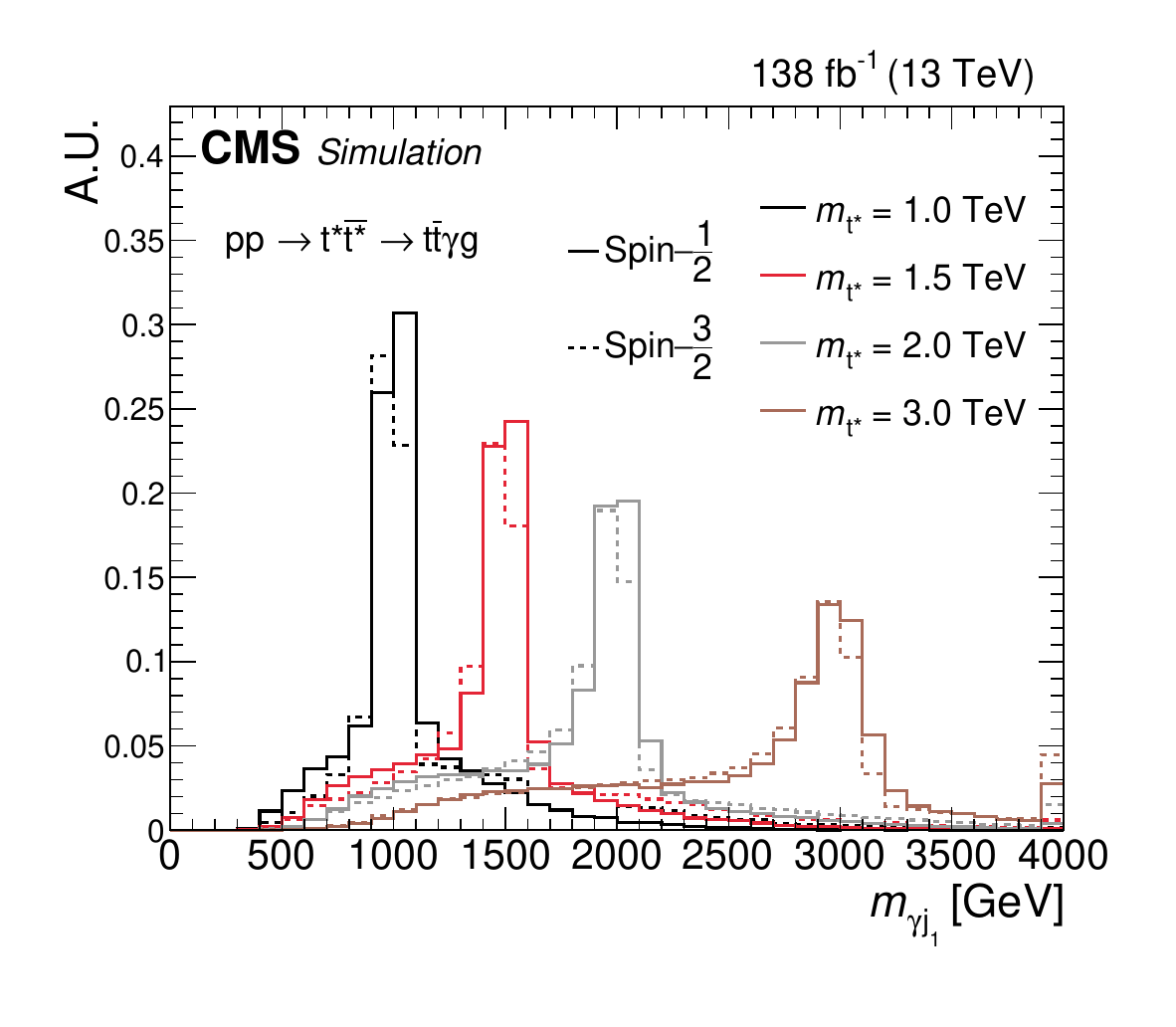}  
  \caption{
  Shape comparison of the \mgj  distribution for \tstartstar signal events generated at different  \mtstar, for the spin-\onehalf (solid) and spin-\threehalf (dashed) cases.
  }
  \label{mtj}
\end{figure}

Figure~\ref{mtj} shows the  \mgj  distribution for \tstartstar signal events in the SR, for different \mtstar.
The peak shape is similar for both the spin-\onehalf and spin-\threehalf signals.

\section{Background estimation}
\label{sec:bg}

After applying the selection criteria, the \Gjets background is substantially reduced by requiring a \PQt-tagged AK8 jet, yet it still accounts for approximately 45\% of the remaining total background. 
The \ttG process constitutes roughly 50\%, with all remaining backgrounds contributing less than 5\% in total.
Background events are categorized based on their photon origin into three groups:
\begin{itemize}
    \item events with a prompt photon, mostly from the \Gjets and \ttG processes;
    \item events with hadrons misidentified as photons, mostly from  fully hadronic \ttbar decays, referred to hereafter as ``\EHfakes'';
    \item events with electrons misidentified as photons, mostly from  \ttbar events in the lepton+jets channel,  referred to hereafter as ``\EGfakes''.
\end{itemize} 
In simulation, a reconstructed photon is classified as ``prompt''
if it is matched to a radiated photon from a primary lepton, quark, or heavy boson. If
the selected  photon is matched to an electron, it is classified as ``\EGfakes''. 
If it comes from other types of particles (\eg, photons from $\PGpz \to \PGg \PGg$
decays or any other hadrons), or no match is found (\eg, photons from pileup interactions), it is classified as ``\EHfakes''.

Non-top quark jets, defined as AK8 jets not originating from the hadronic decays of a top quark, can be misidentified as \PQt-tagged jets, and thus pass the SR selection criteria.
Among background events with non-top quark AK8 jets in the SR, the \Gjets process is dominant, contributing more than $99\%$ of events. Therefore, we hereafter call the sum of all background processes  with a prompt photon and non-top quark AK8 jets the \Gjets background.
Detector simulations cannot accurately replicate the top quark mistag rate observed in data; this mistag rate depends not only on the detector performance, but also on the event topology. 
Additionally, the  \Gjets  sample is generated at LO, and precise higher-order cross section calculations for the restricted phase space of our analysis are not available.
Therefore, we estimate the contribution of the \Gjets background directly from data. 
Similarly, the \EHfakes background is not precisely simulated in the MC-generated samples.
Thus, we also estimate the \EHfakes background from data, although it is a small background. Contributions of other processes are estimated from simulation.

The \Gjets background is estimated with the following strategy. First, the probability of an AK8 jet not originating from a top quark to be \PQt tagged, which is  called the top quark mistag rate ($r$), is measured. Then, the contribution of the \Gjets background in the SR is estimated using data events in a \Gjets application region (AR). The \Gjets AR is defined by requiring the same selection as the SR except for the number of \PQt-tagged jets, which must be equal to zero. The mistag rate $r$ depends only on the \pt and \msd of the jet.
The probability for an event without any top quarks, containing  $n$  AK8 jets, to have exactly zero \PQt-tagged jets is expressed as:
\begin{equation}
P = P_{\text{zero \PQt-tagged jets}}(\pt, \msd)=\prod_{i=1}^n (1-r_i(\pt,\msd)).
\end{equation}
The \Gjets background in the SR is estimated by applying a weight of  $(1-P)/P$  to events in the AR.

The top quark mistag rate $r$ is measured using data events in a  region with low signal contamination defined by the presence of one photon that passes a looser photon ID condition compared with the medium ID criteria but fails the medium ID selection, together with an AK8 jet. 
This region is enriched with multijet events featuring a similar proportion of \PQb and \PQc quark jets, as well as the combined contribution from jets originating from light-flavor quarks and gluons (\PQu, \PQd, \PQs quarks and \Pg), as seen in \Gjets events in simulation.
 The mistag rate $r$ is defined as the ratio of the number of events in which the AK8 jet is \PQt-tagged to the total number of selected events.  The contribution from processes involving top quarks is subtracted from the data using simulated samples, accounting for less than 10\% of events with one \PQt-tagged AK8 jet. The mistag ratio $r$ is extracted in bins of AK8 jet \pt and \msd.

The \Gjets background estimation is validated in a region with a photon and exactly one  AK8 jet that is identified as \PQt tagged. We divide this validation region into two subregions based on the \msdj: $125 < \msdj < 225$\GeV (VR1), similar to the SR, and $\msdj < 125$\GeV or $\msdj > 225$\GeV (VR2). Both VR1 and VR2 have a large \Gjets contribution and minimal signal contamination. The VR2 is used to perform a closure test of the \Gjets estimation, whereas VR1 is included in the final fit to constrain the \Gjets background. 

The \EHfakes background consists of fully hadronic \ttbar decay events, where energy deposits from hadrons in the detector are misidentified as originating from a prompt photon. These hadrons tend to have abnormally high fractions of energy deposited in the ECAL.
This background is estimated from data using  the misidentification rate method~\cite{CMS:2021ctt}.
To use the method, we measure the probability (\fr) that a nonprompt photon passes the medium photon ID selection criteria in a multijet sample. 
The value \fr is defined as the ratio
of the number of events with a nonprompt photon candidate where the photon passes the medium criteria
to the total number of events. 
The ``ABCD'' method, a background estimation technique using data that relies on factorizing signal and background contributions in independent control regions (CRs), is used to measure the numerator of \fr~\cite{ATLAS:2023qdu}.
Since the \sii and \Ich variables are almost uncorrelated, three sideband CRs are defined by inverting medium ID requirements on the \sii and \Ich variables. 
The fraction of events with a nonprompt photon candidate in the CRs is larger than 98\%.
The denominator of \fr is calculated by adding the number of data events in which the photon fails the medium ID to the numerator of \fr.
The contribution of events with a prompt photon that fails the medium ID is subtracted from data  using MC predictions from simulation. The value of \fr is measured in bins of the reconstructed photon's $\eta$.    
The \EHfakes background contributions in the SR, VR1, and VR2 are estimated by reweighting data events that satisfy the same selection criteria, except that the photon is required to fail the medium ID criteria. Each event is assigned a weight of $\fr / (1 - \fr)$.

\begin{figure*}[tb]
\centering

\includegraphics[height=0.26\textheight]{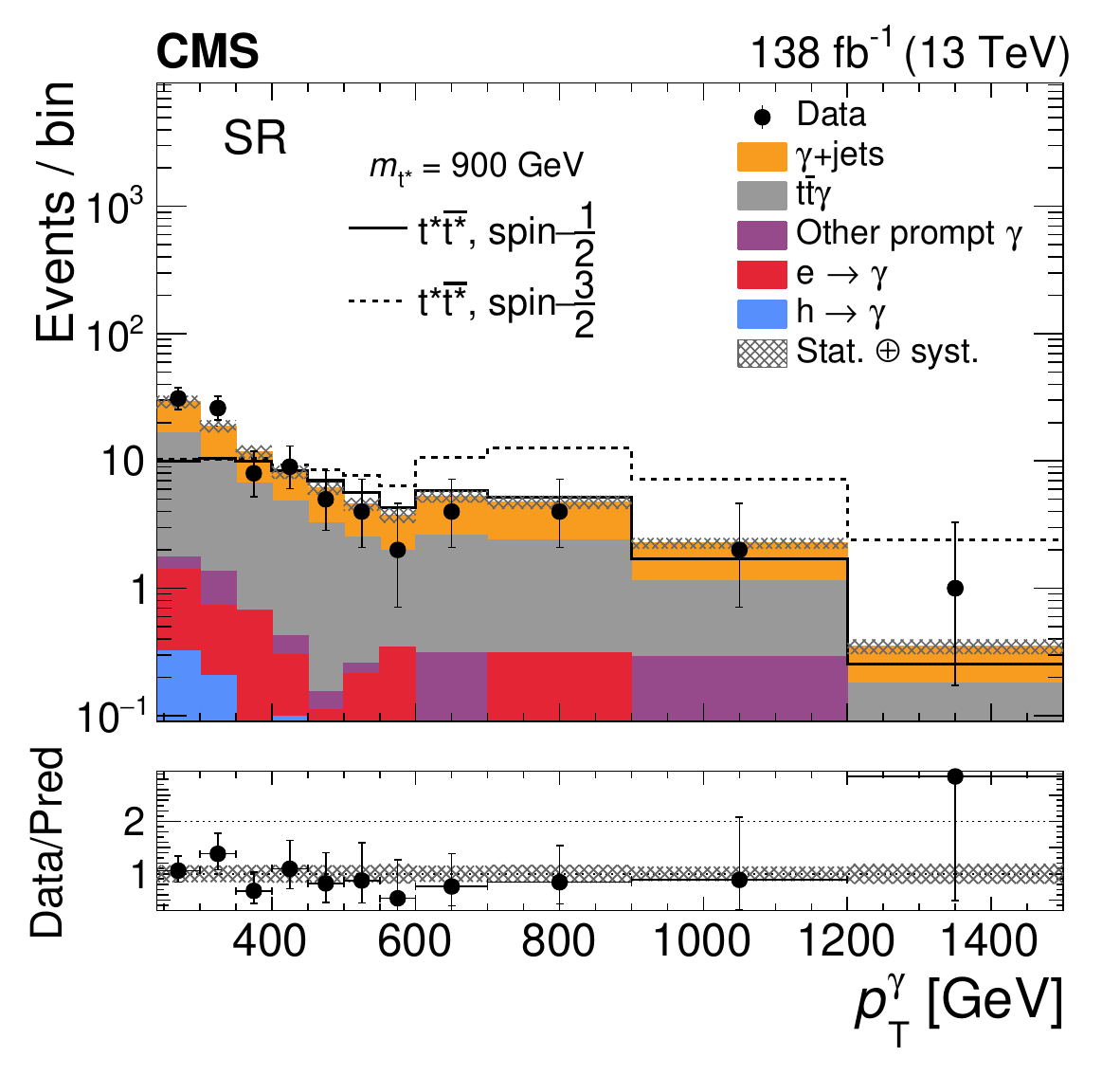}
\includegraphics[height=0.26\textheight]{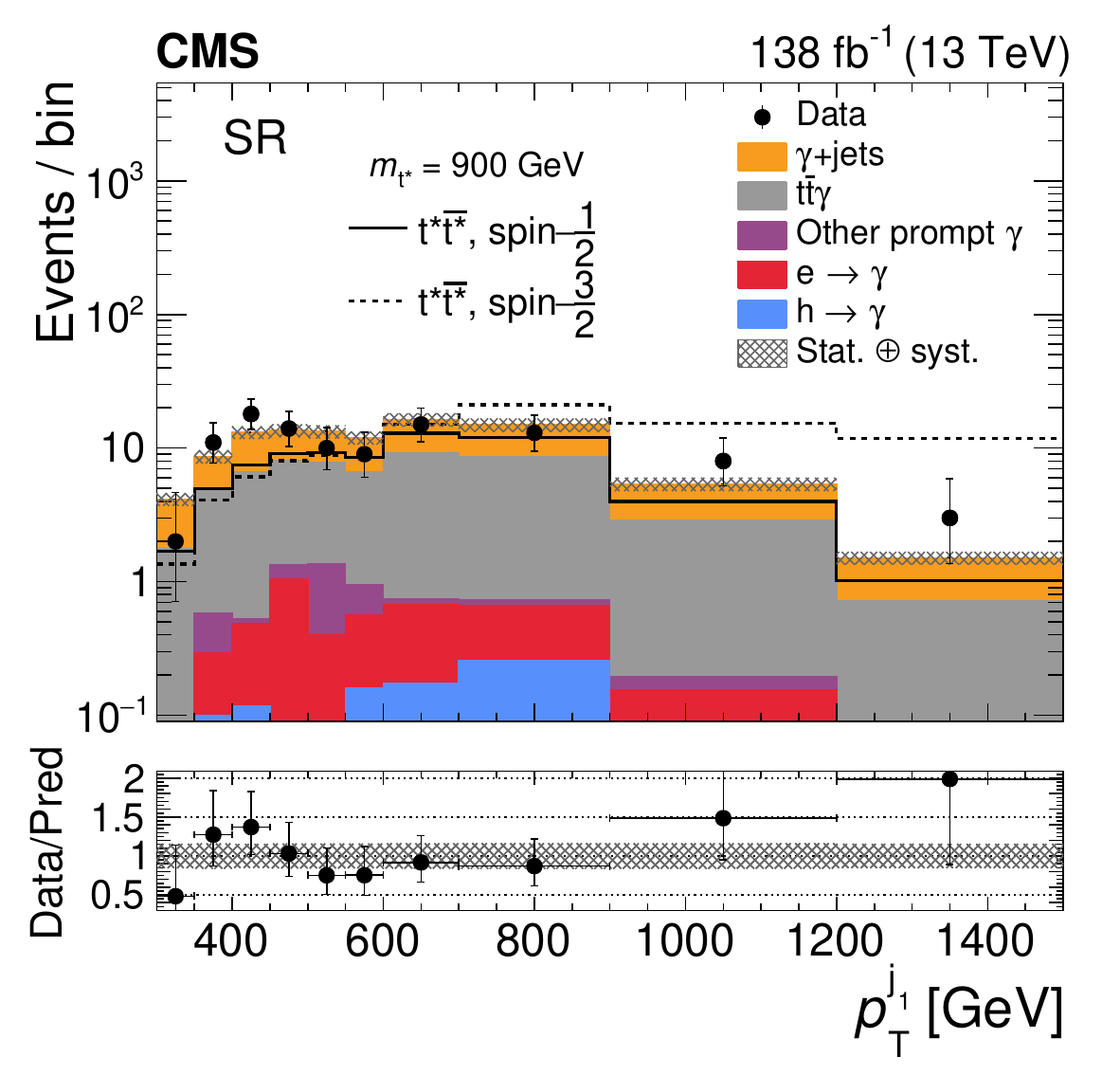}

\includegraphics[height=0.26\textheight]{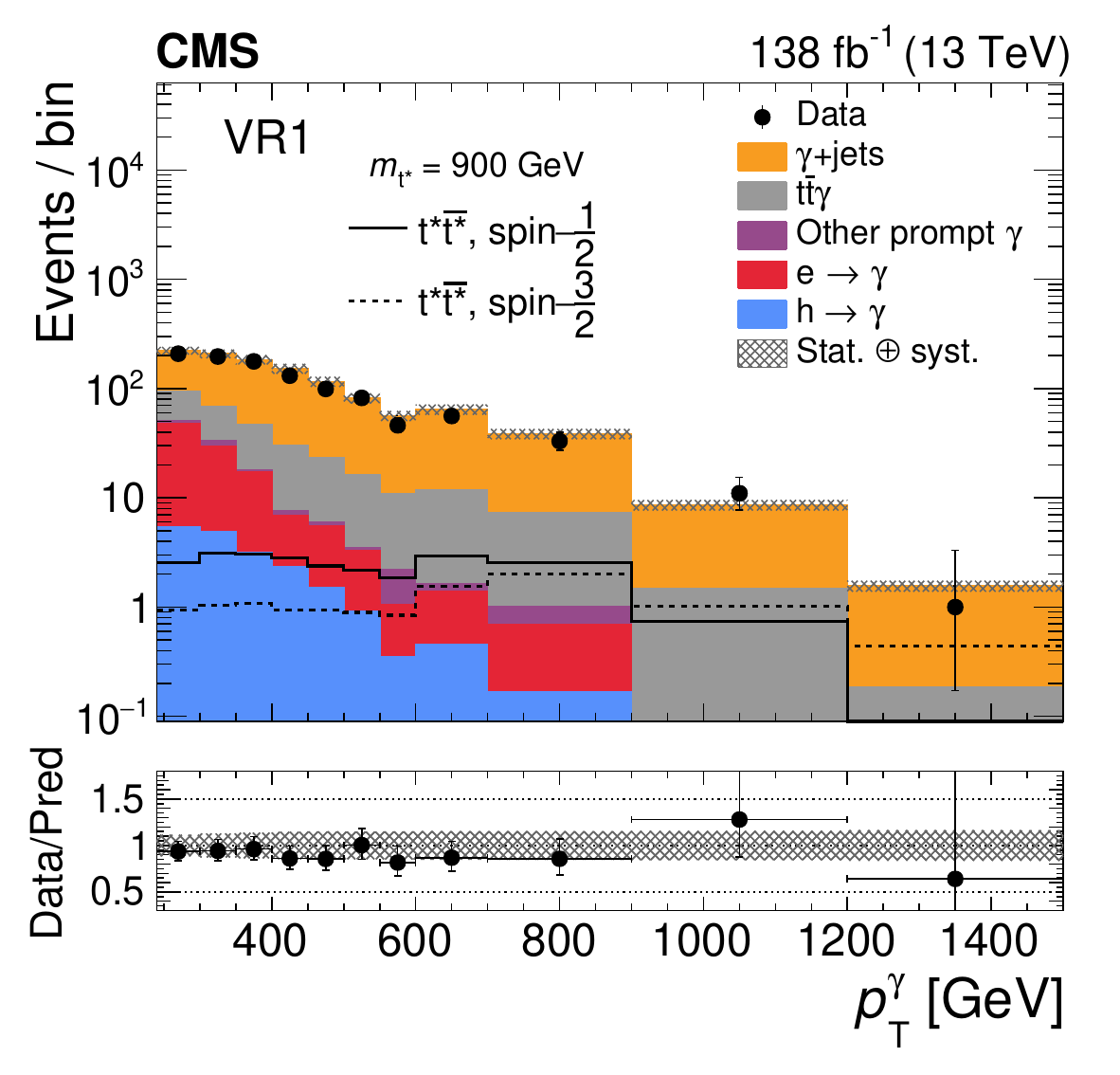}
\includegraphics[height=0.26\textheight]{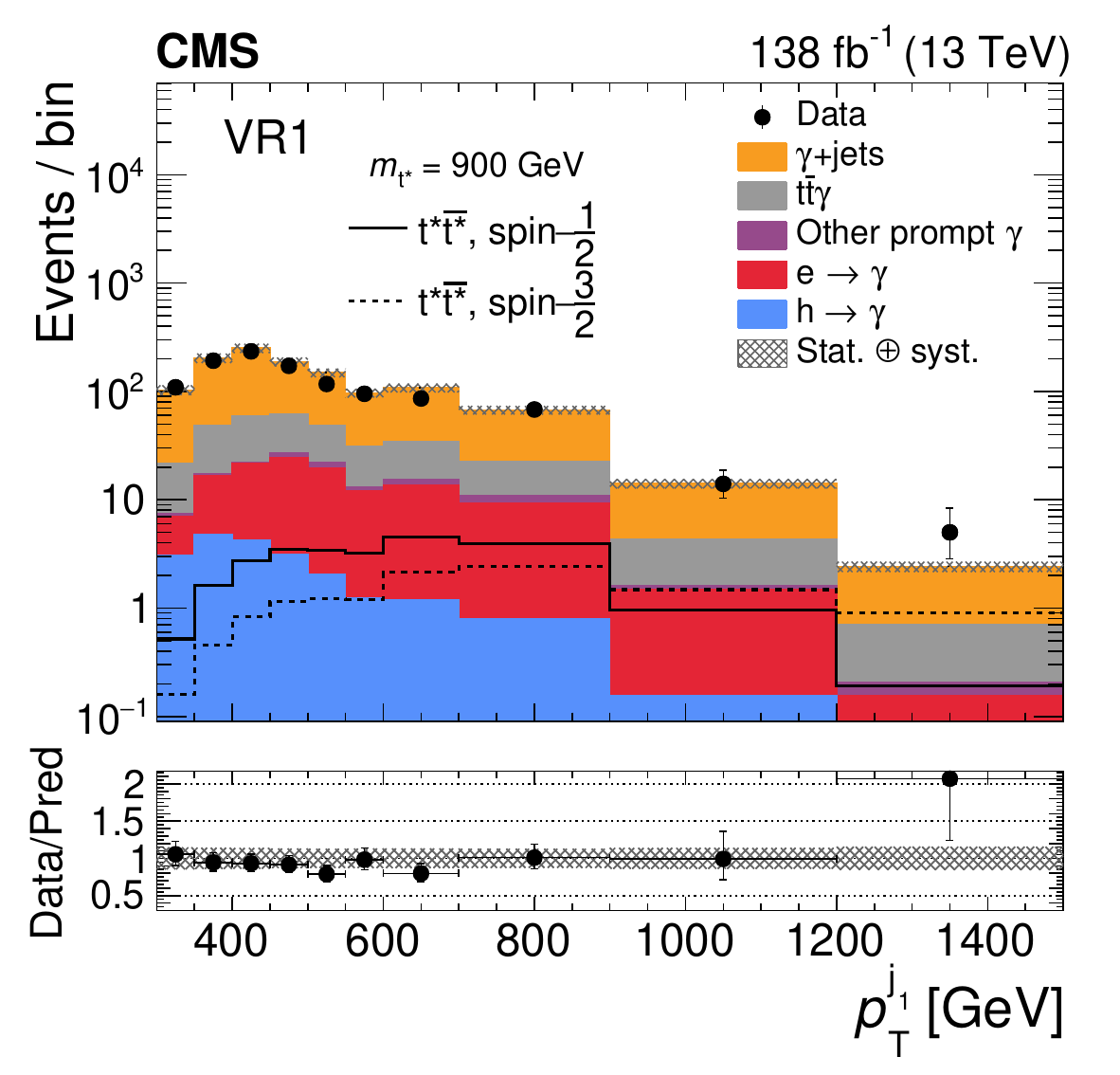}

\includegraphics[height=0.26\textheight]{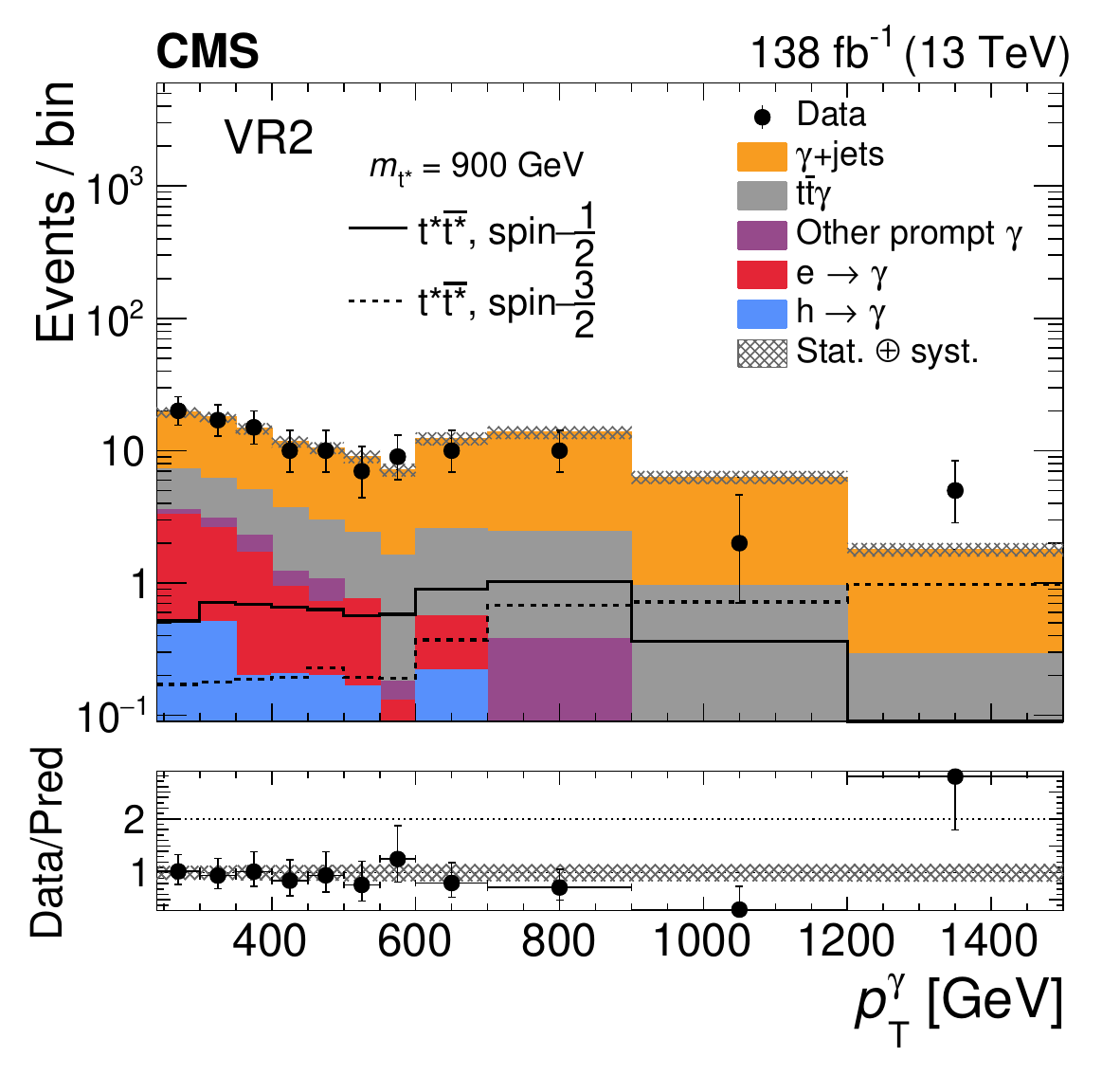}
\includegraphics[height=0.26\textheight]{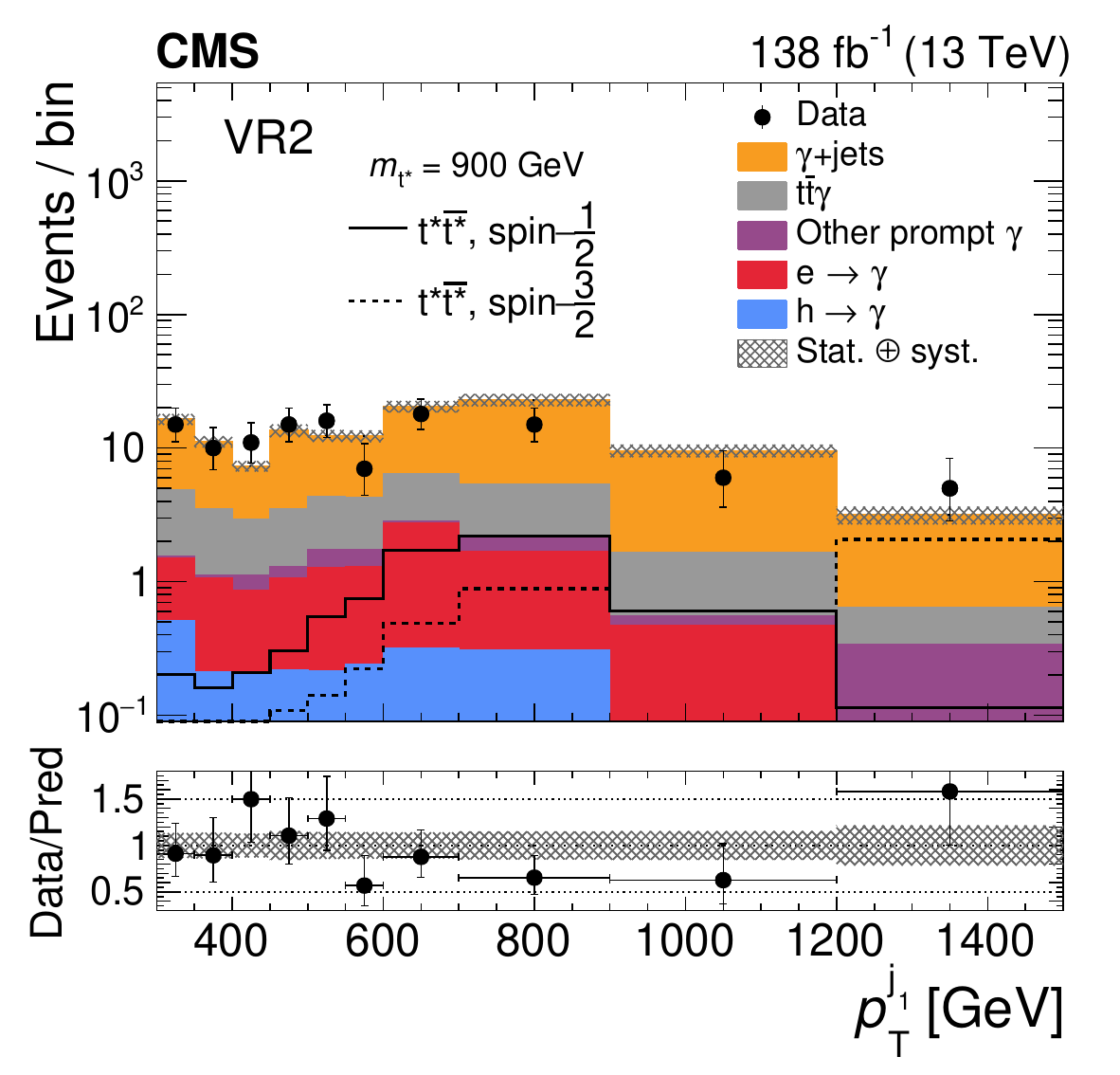}

\caption{The photon (left) and j$_1$ (right) \pt distributions in SR (upper), VR1 (middle), and VR2 (lower) are presented for data and expected backgrounds. Statistical and systematic uncertainties in the expected background yields are depicted by the hatched bands. Additionally, the simulated signal distributions for spin-\onehalf and spin-\threehalf $\tstar$ with a mass of 900\GeV are overlaid for comparison, with both samples normalized to a cross section of 10\unit{fb}. The last bins include the overflow. The lower panels show the ratio of the data to the total SM background, with the total uncertainties indicated by the shaded gray band.}
\label{kin1}
\end{figure*}

The contributions from \Gjets and \EHfakes backgrounds are derived  from data, and other backgrounds, including $\ttG$ and \EGfakes, are estimated from simulations. All backgrounds containing a prompt photon alongside a $\PQt$, $\PW$, or $\PZ$, except for $\ttG$, are grouped together and labeled as ``other prompt \PGg''.
Figure~\ref{kin1}  shows the distributions of $\pt$  for \PGg and j$_1$ in SR, VR1, and VR2. The estimation of the \Gjets background is in good agreement with the observed data in both VR1 and VR2. Additional kinematic distributions in VR1 and VR2 are examined to validate the mistag-rate method and to check for possible residual dependencies on other event variables.

\section{Systematic uncertainties}
\label{sec:uncertainties}
Systematic uncertainties, both experimental and theoretical, influence the predicted background yields, distributions, and signal selection efficiency.  Uncertainties affecting only event yields are modeled using log-normal distributions, whereas those that impact the shape of distributions are represented with Gaussian distributions. This dual modeling approach effectively corrects for variations in both normalization and kinematic distributions.

 The integrated luminosities for the 2016, 2017, and 2018 data-taking periods carry uncertainties of 1.2--2.5\%~\cite{CMS-LUM-17-003,CMS-PAS-LUM-17-004,CMS-PAS-LUM-18-002}, and the combined uncertainty for the entire 2016--2018 period is 1.6\%. Pileup modeling uncertainties are evaluated by varying the total inelastic pp cross section by $\pm$4.6\%~\cite{CMS:2018mlc}, which directly affects the pileup weight applied in simulation.

 Trigger SFs are derived from the ``tag-and-probe'' method \cite{CMS:2011aa}, using $\PZ\to\Pe\Pe$ events, to correct the simulated event yields. The associated systematic uncertainties, extracted from the same method,  are  1\% for photons with $\pt >240\GeV$. 
 
The systematic uncertainties associated with photon identification SFs are evaluated by varying these SFs within their respective uncertainties. Additional uncertainties in the photon energy scale and resolution are quantified using $\PZ\to\Pe\Pe$ events, treating the electrons as photon-like by ignoring the track information of the electrons. These events provide a reliable control sample to study discrepancies between measured and simulated photon properties. All photon-related uncertainties, including those for trigger, identification, energy scale, and resolution, are treated as fully correlated across the three data-taking years.

Uncertainties in the jet energy scale and resolution are evaluated by varying the \pt of the reconstructed jets in simulated events separately for each of the various uncertainty sources described in Ref.~\cite{CMS:2016lmd}. These uncertainty components are treated based on their origin; some are considered fully correlated across the  data-taking eras (\eg, uncertainties from common calibration procedures), whereas others, such as those arising from year-specific detector conditions or from a limited event count, are treated as uncorrelated.
For the \msd scale, uncertainties are quantified using the observed shift in the W boson  peak mass in a sample enriched with $\ttbar$ events in the lepton+jets channel~\cite{CMS-DP-2023-044}.
Efficiency corrections for the \PQt-tagging working point are derived from \ttbar events in the leptons+jets channel and applied as a function of AK8 jet $\pt$ and $\abs{\eta}$ for AK8 jets originating from top quarks. Systematic uncertainties are assessed by varying the correction factors within their uncertainties. Uncertainties associated with the \msd scale and top quark tagging SF are treated as correlated across the data-taking years.

Theoretical uncertainties due to the renormalization  and  factorization scales in the matrix element generators are modeled by varying the scales independently by a factor of 0.5 or 2. 
We exclude the extreme variations  (i.e., variations of the two scales in the opposite directions), ensuring that only the relevant variations are included in the uncertainty estimation.
The PDF uncertainty is evaluated using a set of replicas of NNPDF3.1~\cite{Ball:2017nwa}.
The final PDF uncertainty is calculated by evaluating distributions for each replica and determining the spread in the predictions. The standard deviation of the results across all replicas quantifies the PDF uncertainty.
These theoretical uncertainties are considered for only the signal and \ttG samples and are implemented by including their modifications to the shapes of kinematic distributions that affect the experimental detection efficiency and signal acceptance.
A conservative 30\% uncertainty is applied to the normalization of \EGfakes and other prompt \PGg  backgrounds, which contribute ~2\% and ~3\% to the total background in the SR, respectively~\cite{CMS:2022lmh}.
All sources of theoretical uncertainties are treated as fully correlated across the data-taking years. 

The primary source of uncertainty in the \Gjets background estimate arises from the statistical uncertainty in the transfer function $r(\pt, \msd)$. Each bin of $r(\pt, \msd)$ is treated as an independent source of uncertainty, allowing for a granular assessment. Additionally, a 15\% uncertainty on the overall normalization of the \Gjets background is included to account for systematic variations. This value is consistent with the level of agreement observed in VR2 between the data and the prediction.
The contribution from the \EHfakes background to the SR is negligible ($<1\%$), but a conservative uncertainty of 50\% is assigned to this component to account for possible variations. Varying this uncertainty by a factor of two has no noticeable impact on the results.

\section{Results}

The statistical interpretation of the results in this analysis is carried out using the CMS  \textsc{Combine} framework~\cite{CMS:2024onh}. A maximum likelihood fit is applied to the \mgj distributions in the SR and VR1, with the systematic uncertainties discussed in Section \ref{sec:uncertainties} incorporated as nuisance parameters.
To address the finite sizes of the simulated samples, each bin of the simulated signal-plus-background template is varied independently within its statistical uncertainty. This is done following the Barlow--Beeston ``lite'' approach~\cite{bbb}.
While including VR1 in the fit aids in constraining the \Gjets background, adding VR2 does not improve the results due to its low statistical significance.
Therefore, VR2 is used solely to validate the \Gjets background and is not included in the final results.

\begin{figure*}[tb]
\centering
       \includegraphics[width=0.4\textwidth]{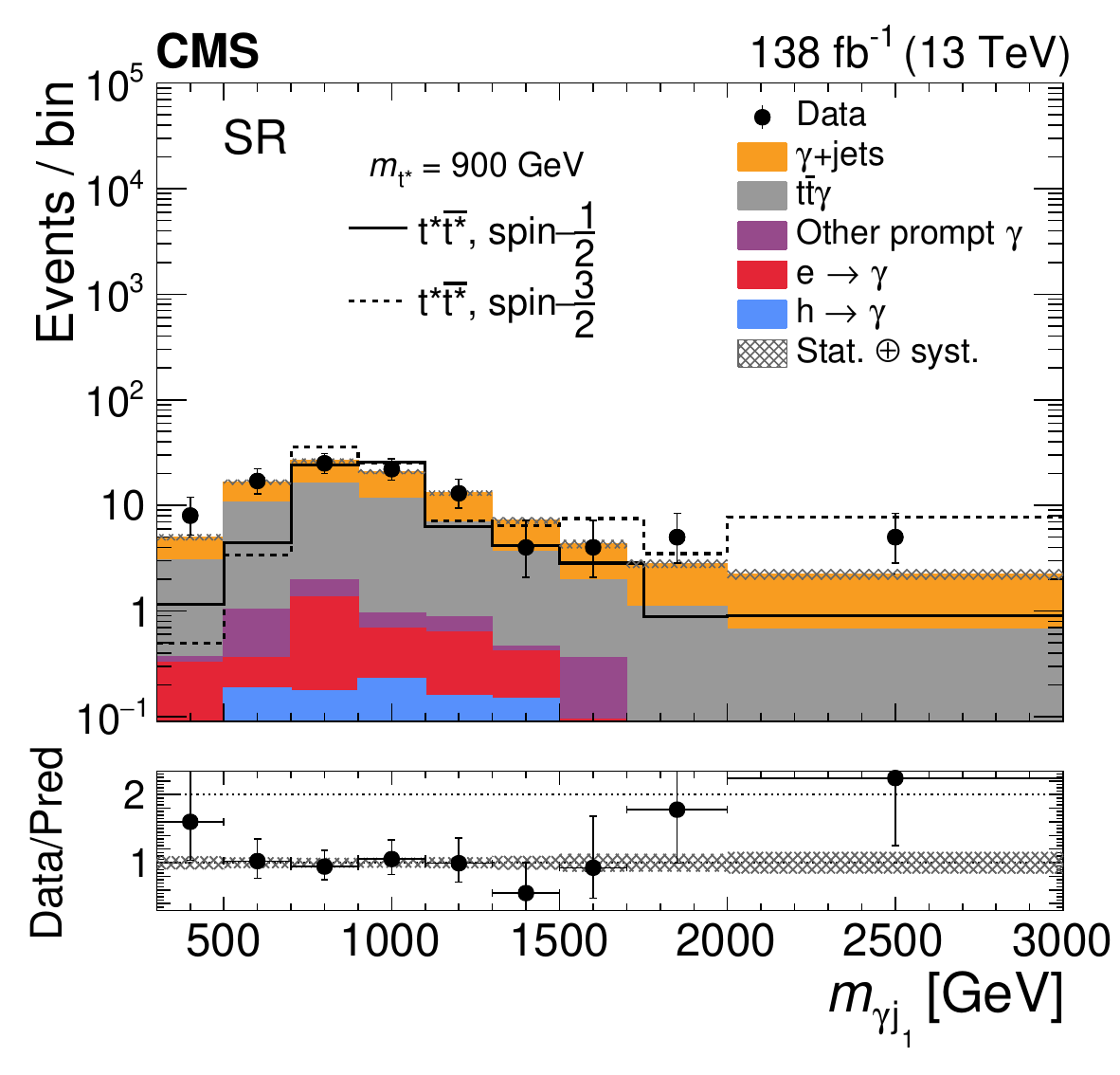}
            \includegraphics[width=0.4\textwidth]{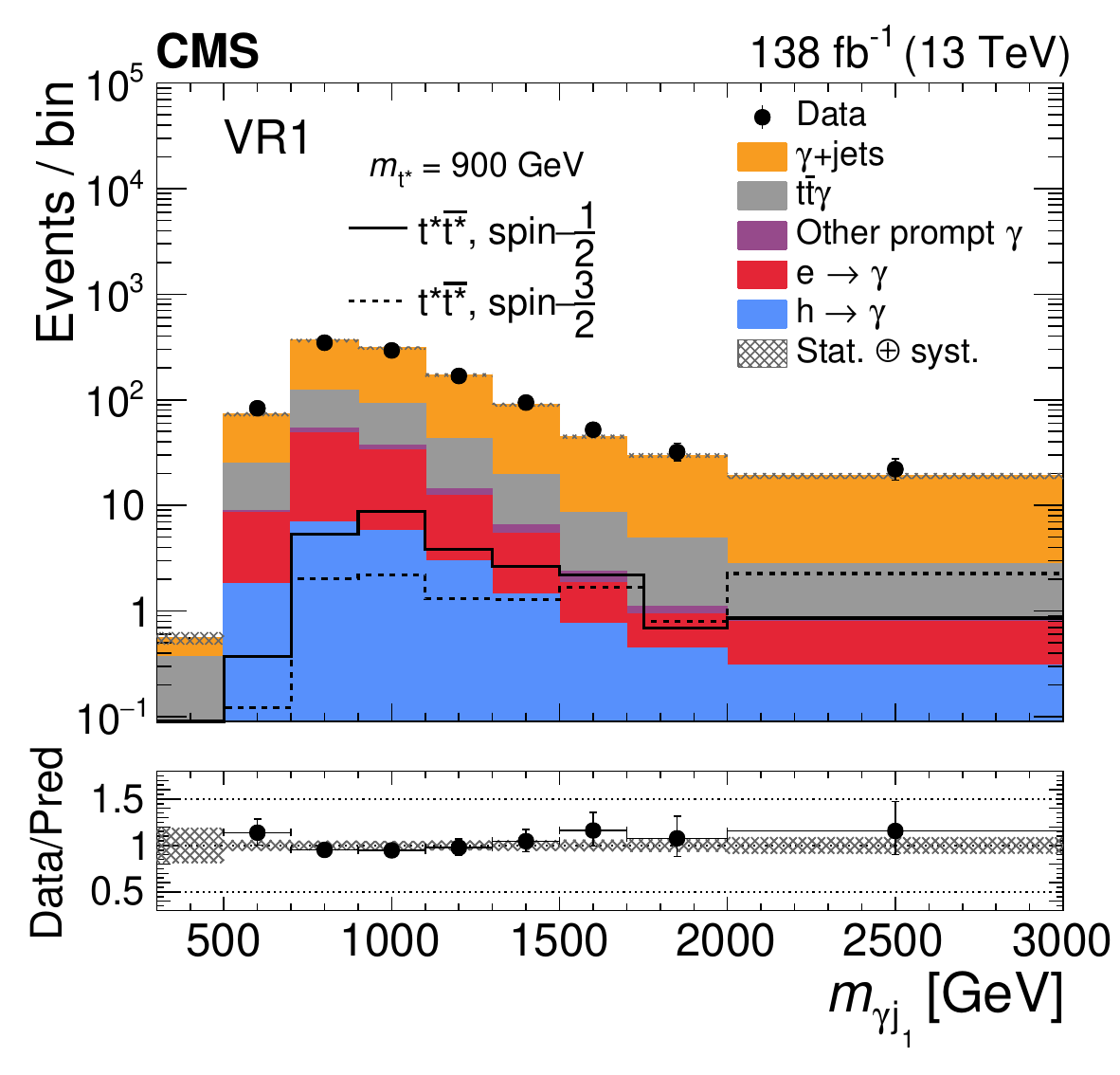}
  \caption{  The background-only postfit $\mgj$ distributions for the SR (left) and VR1 (right) are shown.  Notations are as in Fig.~\ref{kin1}.}
  \label{prefit2}
\end{figure*}

The background-only fits to the reconstructed \mtstar distributions are shown in Fig.~\ref{prefit2} for the SR and VR1. 
The  \Gjets background describes the data in VR1 within uncertainties.
The binning of the \mgj distribution is chosen to ensure a sufficient number of data events in the $\Gjets$ AR, while preserving sensitivity to a potential signal resonance.
No notable excess is observed at low masses in the SR. However, at high masses, the data exhibit a deviation from the background expectation, reaching a maximum local significance of approximately  $~2.5$ standard deviations.  After accounting for the look-elsewhere effect over the full mass range probed, the global significance is reduced to approximately $1.8$ standard deviations. 
Variations of the fit configuration, including changes to the treatment of the \Gjets background uncertainties, have a negligible impact on the observed significance.
 In addition, distributions for signal samples with $\mtstar=900\GeV$ are shown for both spin-\onehalf (solid lines) 
 and spin-\threehalf \tstar (dotted lines). 
Both signal samples exhibit distinct peaks at 900\GeV; however, the spin-\threehalf signal features a more pronounced tail extending toward higher \mtstar, whereas the spin-\onehalf signal shows a broader distribution toward lower \mtstar. 

\begin{figure*}[tb]
\centering
\includegraphics[width=0.4\textwidth, clip]{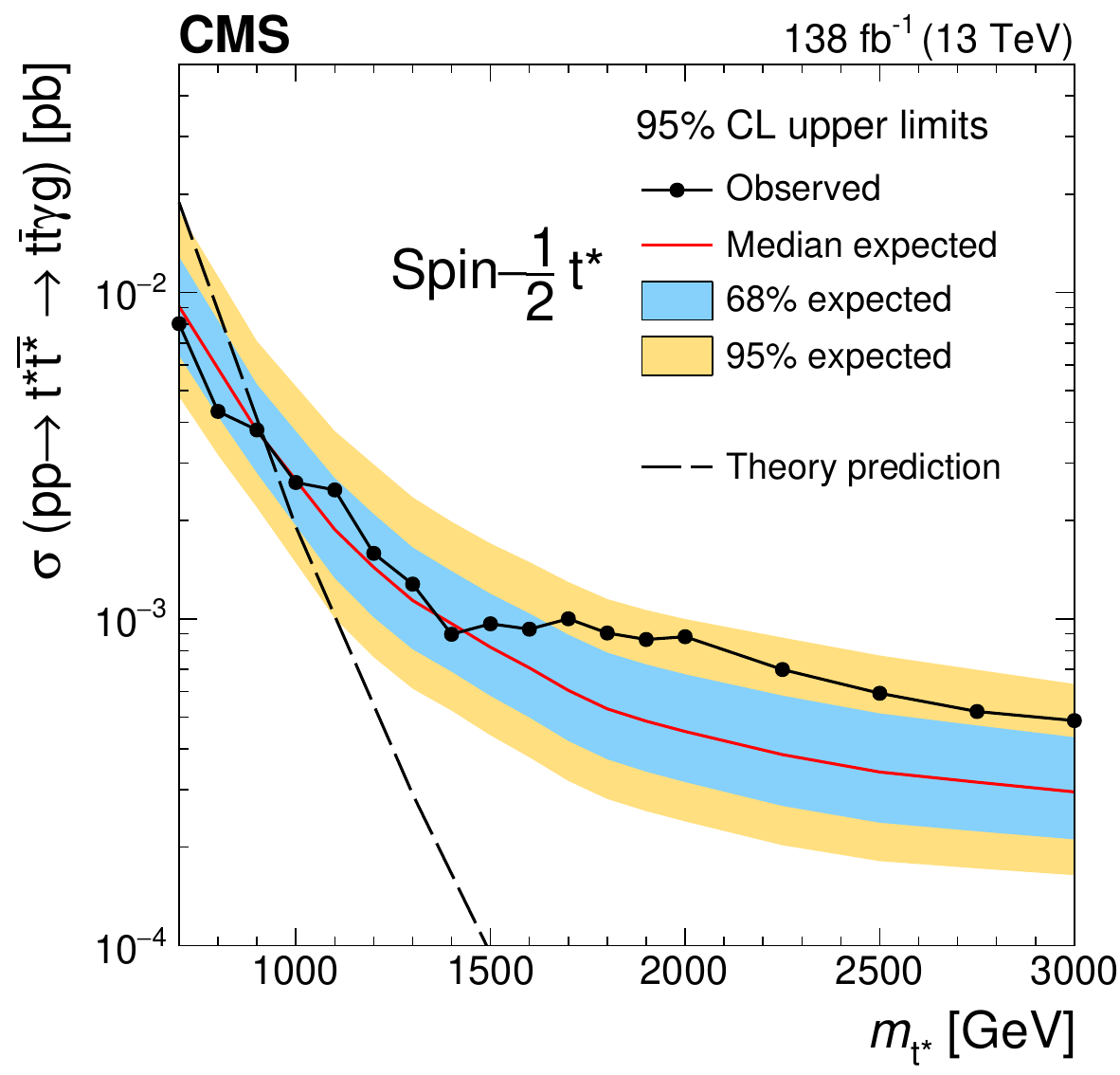} 
\includegraphics[width=0.4\textwidth, clip]{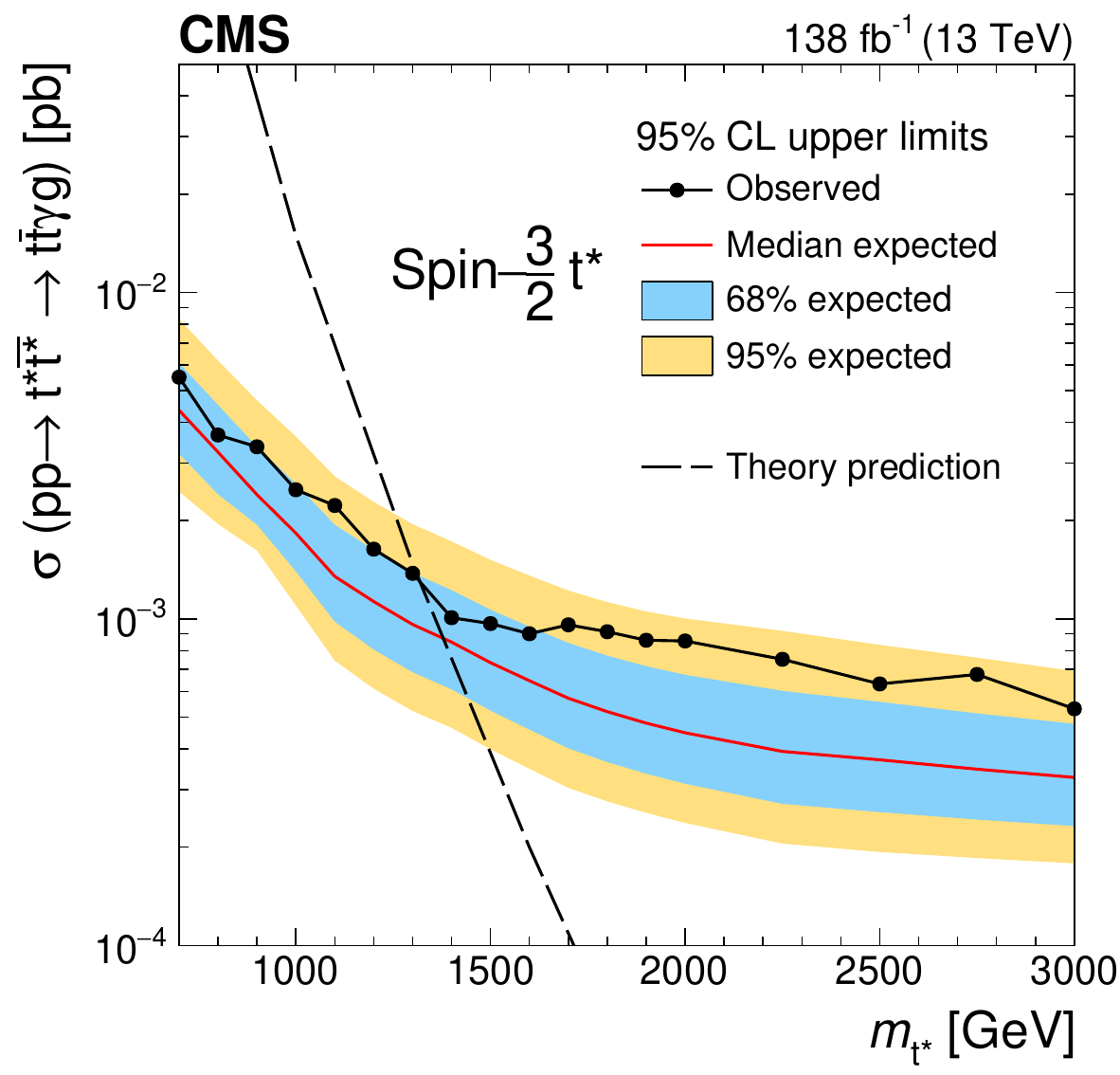} 
\caption{The expected (red solid lines) and the observed (black solid lines) 95\% \CL upper
limits on the cross section of the $\pptstartstartgta$ process as functions of  the \mtstar for a spin-\onehalf (left) and spin-\threehalf (right) $\tstar$.  The
blue (inner) and yellow (outer) bands indicate the 68\% and 95\% coverage of the expected upper
limits. The black dashed lines indicate the theoretical expectations~\cite{Alhazmi:2018whk}}.
\label{limit}
\end{figure*}

Upper limits on the \tstartstar production cross section times the branching fraction \BR(\tstartstartgta), are determined for both the spin-\onehalf and spin-\threehalf hypotheses. These limits are obtained using the distributions of the likelihood ratio, calculated with the asymptotic approximation~\cite{Cowan:2010js}, and the \CLs criterion~\cite{CMS-NOTE-2011-005}.
The excluded phase space as a function of \mtstar obtained from the expected and observed upper limits at 95\% confidence level (\CL) is shown in Fig.~\ref{limit}.
The three key sources of uncertainty that strongly correlate with the signal strength parameter and are consistent across different \tstar masses are the normalization uncertainty of the \Gjets process, the renormalization and factorization scale uncertainty in $\ttG$ simulation, and the uncertainty in the signal PDF.
The observed (expected) 95\% \CL lower limits on the mass of the spin-\onehalf and spin-\threehalf \tstar are  
\SohLimitO (\SohLimitE)\GeV and \SthLimitO (\SthLimitE)\GeV, respectively.
Although the limits on $\mtstar$ depend on the assumed relative branching fractions of $\tstarta$ and $\tstartg$, we compare our results, assuming $\BR(\tstarta) = 3\%$ and $\BR(\tstartg) = 97\%$, with  the limits obtained in the $\pptstartstartgtg$ search by the CMS Collaboration, which assumes $\BR(\tstartg) = 100\%$~\cite{CMS:2024zpb}, as presented in Table~\ref{tab:tstar_limits}. These limits could vary due to the dependence of the theoretical prediction on the assumed values of $\BR(\tstarta)$ and $\BR(\tstartg)$, even though the $\pptstartstar$ production cross section is identical in both analyses.
Tabulated results are provided in the HEPData record for this analysis~\cite{hepdata}.

\begin{table}[tb]
    \centering
    \caption{Observed and expected 95\% \CL lower mass limits for \tstar searches comparing this analysis to previous results in the \tgtg channel~\cite{CMS:2024zpb}.}
    \begin{scotch}{lcccc}
        \multirow{2}{*}{Channel}  & \multicolumn{2}{c}{ Observed [\GeVns{}]} & \multicolumn{2}{c}{expected [\GeVns{}]} \\
        &  spin-\onehalf & spin-\threehalf & spin-\onehalf & spin-\threehalf \\  
        \hline      
        \pptstartstartgta  & \SohLimitO & \SthLimitO& \SohLimitE & \SthLimitE \\
        \pptstartstartgtg  & 1050 & 1700 & 990 & 1690 \\
    \end{scotch}
    \label{tab:tstar_limits}
\end{table}
\clearpage
\section{Summary}
A search has been presented for spin-\onehalf and spin-\threehalf heavy particles in a final state with a photon and large-radius jets. The data
correspond to an integrated luminosity of 138\fbinv collected between 2016 and 2018
with the CMS detector at the LHC in proton-proton collisions at $\sqrt{s}=13\TeV$. 
The pair production of excited top quarks is searched for, considering the decay to (top quark and antiquark, photon and gluon) signature \tstartstartgta.
Lorentz-boosted top quark identification techniques are used to select  events containing
large-radius jets originating from top quarks and to suppress standard model (SM) backgrounds.
A small deviation from the SM expectations with a local significance of 2.5 standard deviations is seen in the data.  Upper limits are placed on the product of
the \tstartstar  cross section and $\mathcal{B}(\tstartstartgta)$, as a
function of the \tstar mass.
A spin-\onehalf (spin-\threehalf) \tstar with a mass below \SohLimitO (\SthLimitO)\GeV is excluded at 95\% confidence level.
These are the first limits obtained from  the \tstartstartgta channel. They are competitive to the limits obtained from  the \tstartstartgtg channel for a spin-\onehalf\tstar using complementary techniques.

\begin{acknowledgments}
We congratulate our colleagues in the CERN accelerator departments for the excellent performance of the LHC and thank the technical and administrative staffs at CERN and at other CMS institutes for their contributions to the success of the CMS effort. In addition, we gratefully acknowledge the computing centers and personnel of the Worldwide LHC Computing Grid and other centers for delivering so effectively the computing infrastructure essential to our analyses. Finally, we acknowledge the enduring support for the construction and operation of the LHC, the CMS detector, and the supporting computing infrastructure provided by the following funding agencies: SC (Armenia), BMBWF and FWF (Austria); FNRS and FWO (Belgium); CNPq, CAPES, FAPERJ, FAPERGS, and FAPESP (Brazil); MES and BNSF (Bulgaria); CERN; CAS, MoST, and NSFC (China); MINCIENCIAS (Colombia); MSES and CSF (Croatia); RIF (Cyprus); SENESCYT (Ecuador); ERC PRG and PSG, TARISTU24-TK10 and MoER TK202 (Estonia); Academy of Finland, MEC, and HIP (Finland); CEA and CNRS/IN2P3 (France); SRNSF (Georgia); BMFTR, DFG, and HGF (Germany); GSRI (Greece); NKFIH (Hungary); DAE and DST (India); IPM (Iran); SFI (Ireland); INFN (Italy); MSIT and NRF (Republic of Korea); MES (Latvia); LMTLT (Lithuania); MOE and UM (Malaysia); BUAP, CINVESTAV, CONACYT, LNS, SEP, and UASLP-FAI (Mexico); MOS (Montenegro); MBIE (New Zealand); PAEC (Pakistan); MES, NSC, and NAWA (Poland); FCT (Portugal);  MESTD (Serbia); MICIU/AEI and PCTI (Spain); MOSTR (Sri Lanka); Swiss Funding Agencies (Switzerland); MST (Taipei); MHESI (Thailand); TUBITAK and TENMAK (T\"{u}rkiye); NASU (Ukraine); STFC (United Kingdom); DOE and NSF (USA).

\hyphenation{Rachada-pisek} Individuals have received support from the Marie-Curie program and the European Research Council and Horizon 2020 Grant, contract Nos.\ 675440, 724704, 752730, 758316, 765710, 824093, 101115353, 101002207, 101001205, and COST Action CA16108 (European Union); the Leventis Foundation; the Alfred P.\ Sloan Foundation; the Alexander von Humboldt Foundation; the Science Committee, project no. 22rl-037 (Armenia); the Fonds pour la Formation \`a la Recherche dans l'Industrie et dans l'Agriculture (FRIA) and Fonds voor Wetenschappelijk Onderzoek contract No. 1228724N (Belgium); the Beijing Municipal Science \& Technology Commission, No. Z191100007219010, the Fundamental Research Funds for the Central Universities, the Ministry of Science and Technology of China under Grant No. 2023YFA1605804, the Natural Science Foundation of China under Grant No. 12535004, and USTC Research Funds of the Double First-Class Initiative No.\ YD2030002017 (China); the Ministry of Education, Youth and Sports (MEYS) of the Czech Republic; the Shota Rustaveli National Science Foundation, grant FR-22-985 (Georgia); the Deutsche Forschungsgemeinschaft (DFG), among others, under Germany's Excellence Strategy -- EXC 2121 ``Quantum Universe" -- 390833306, and under project number 400140256 - GRK2497; the Hellenic Foundation for Research and Innovation (HFRI), Project Number 2288 (Greece); the Hungarian Academy of Sciences, the New National Excellence Program - \'UNKP, the NKFIH research grants K 131991, K 133046, K 138136, K 143460, K 143477, K 146913, K 146914, K 147048, 2020-2.2.1-ED-2021-00181, TKP2021-NKTA-64, and 2025-1.1.5-NEMZ\_KI-2025-00004 (Hungary); the Council of Science and Industrial Research, India; ICSC -- National Research Center for High Performance Computing, Big Data and Quantum Computing, FAIR -- Future Artificial Intelligence Research, and CUP I53D23001070006 (Mission 4 Component 1), funded by the NextGenerationEU program (Italy); the Latvian Council of Science; the Ministry of Education and Science, project no. 2022/WK/14, and the National Science Center, contracts Opus 2021/41/B/ST2/01369, 2021/43/B/ST2/01552, 2023/49/B/ST2/03273, and the NAWA contract BPN/PPO/2021/1/00011 (Poland); the Funda\c{c}\~ao para a Ci\^encia e a Tecnologia (Portugal); the National Priorities Research Program by Qatar National Research Fund; MICIU/AEI/10.13039/501100011033, ERDF/EU, ``European Union NextGenerationEU/PRTR", and Programa Severo Ochoa del Principado de Asturias (Spain); the Chulalongkorn Academic into Its 2nd Century Project Advancement Project, the National Science, Research and Innovation Fund program IND\_FF\_68\_369\_2300\_097, and the Program Management Unit for Human Resources \& Institutional Development, Research and Innovation, grant B39G680009 (Thailand); the Eric \& Wendy Schmidt Fund for Strategic Innovation through the CERN Next Generation Triggers project under grant agreement number SIF-2023-004; the Kavli Foundation; the Nvidia Corporation; the SuperMicro Corporation; the Welch Foundation, contract C-1845; and the Weston Havens Foundation (USA).
\end{acknowledgments}

\bibliography{auto_generated} 

\cleardoublepage \appendix\section{The CMS Collaboration \label{app:collab}}\begin{sloppypar}\hyphenpenalty=5000\widowpenalty=500\clubpenalty=5000\cmsinstitute{Yerevan Physics Institute, Yerevan, Armenia}
{\tolerance=6000
A.~Hayrapetyan, V.~Makarenko\cmsorcid{0000-0002-8406-8605}, A.~Tumasyan\cmsAuthorMark{1}\cmsorcid{0009-0000-0684-6742}
\par}
\cmsinstitute{Institut f\"{u}r Hochenergiephysik, Vienna, Austria}
{\tolerance=6000
W.~Adam\cmsorcid{0000-0001-9099-4341}, J.W.~Andrejkovic, L.~Benato\cmsorcid{0000-0001-5135-7489}, T.~Bergauer\cmsorcid{0000-0002-5786-0293}, M.~Dragicevic\cmsorcid{0000-0003-1967-6783}, C.~Giordano\cmsorcid{0000-0001-6317-2481}, P.S.~Hussain\cmsorcid{0000-0002-4825-5278}, M.~Jeitler\cmsAuthorMark{2}\cmsorcid{0000-0002-5141-9560}, N.~Krammer\cmsorcid{0000-0002-0548-0985}, A.~Li\cmsorcid{0000-0002-4547-116X}, D.~Liko\cmsorcid{0000-0002-3380-473X}, M.~Matthewman, I.~Mikulec\cmsorcid{0000-0003-0385-2746}, J.~Schieck\cmsAuthorMark{2}\cmsorcid{0000-0002-1058-8093}, R.~Sch\"{o}fbeck\cmsAuthorMark{2}\cmsorcid{0000-0002-2332-8784}, D.~Schwarz\cmsorcid{0000-0002-3821-7331}, M.~Shooshtari\cmsorcid{0009-0004-8882-4887}, M.~Sonawane\cmsorcid{0000-0003-0510-7010}, W.~Waltenberger\cmsorcid{0000-0002-6215-7228}, C.-E.~Wulz\cmsAuthorMark{2}\cmsorcid{0000-0001-9226-5812}
% ...
\par}
\vskip\cmsinstskip
\dag:~Deceased\\
$^{1}$Also at Yerevan State University, Yerevan, Armenia\\
$^{2}$Also at TU Wien, Vienna, Austria\\
% ...
\end{sloppypar}
\end{document}